\documentclass{article}[12pt]

\pdfoutput=1
\usepackage{epsfig}
\usepackage{graphicx}
\usepackage{color}
\usepackage{hyperref}
\hypersetup{
    colorlinks,%
    citecolor=magenta,%
    filecolor=magenta,%
    linkcolor=blue,%
    urlcolor=blue
}
\usepackage{dcolumn}
\usepackage{bm}
\usepackage{amsmath}
\usepackage{amssymb}
\usepackage{amsxtra}
\usepackage{wasysym}
\usepackage{kbordermatrix}

\newtheorem{theorem}{Theorem}[section]
\newtheorem{definition}[theorem]{Definition}
\newtheorem{thm}[theorem]{Theorem}
\newtheorem{cor}[theorem]{Corollary}
\newtheorem{lem}[theorem]{Lemma}

\renewcommand{\kbldelim}{(}
\renewcommand{\kbrdelim}{)}

\def\bnprf{\noindent {\bf Proof} \ }
\def\edprf{$_{\Box}$\\ \noindent}
\def\ubar#1{\underbar{$#1$}}

\def\mX{\mathcal{I}}
\def\Xpm{I^{\pm}}
\def\Xp{I^{+}}
\def\Xm{I^{-}}
\def\X0{I^{0}}

\def\mS{\mathcal{X}}
\def\Spm{X^{\pm}}
\def\Sp{X^{+}}
\def\Sm{X^{-}}
\def\S0{X^{0}}
\def\hSpm{\hat{X}^{\pm}}
\def\hSp{\hat{X}^{+}}
\def\hSm{\hat{X}^{-}}
\def\hS0{\hat{X}^{0}}

\def\Sig{S}

\def\de{arc}	
\def\DG{\vec{G}}

\def\DE{A}

\def\mA{\mathcal{A}}
\def\mB{\mathcal{B}}
\def\sm{\sigma}
\def\mkm{\ubar{m}_{k}}
\def\smkm{\scriptsize \ubar{m}_{k}}
\def\mlp{\overline{m}_{l}}

\newsavebox{\bsq}
\savebox{\bsq}(6,6){{\scriptsize $\bullet$}}
\newsavebox{\sq}
\savebox{\sq}(6,6){{\scriptsize $\circ$}}

\title{Split digraphs}

\author{M.~Drew LaMar\footnote{Department of Biology, The College of William and Mary, 2137 Integrated Science Center, Williamsburg VA 23185 ({\tt mdlama@wm.edu})}}

\begin{document}

\maketitle

\begin{abstract}
We generalize the class of split graphs to the directed case and show that these split digraphs can be identified from their degree sequences.  The first degree sequence characterization is an extension of the concept of splittance to directed graphs, while the second characterization says a digraph is split if and only if its degree sequence satisfies one of the Fulkerson inequalities (which determine when an integer-pair sequence is digraphic) with equality.
\end{abstract}

\section{Introduction}

Given a graph $G$, we denote the vertex set by $V(G)$ and the edge set by $E(G)$.  Similarly, for a directed graph $\DG$, we denote the vertex set by $V(\DG)$ and arc set by $A(\DG)$.  A graph $G$ is called {\it split} if the vertex set $V(G)$ can be partitioned into a clique and an independent set, either of which can be empty.  In a clique, there is an edge connecting every pair of vertices, while in an independent set, there are no edges connecting any vertices.  More formally, we have the following definition of a split graph.
\begin{definition}[Split graph \cite{Foldes:1977vj}]
\label{def:splitg}
A graph $G$ is {\bf split} if and only if $V(G)$ is a disjoint union of two sets $A$ and $B$ such that $A$ is a clique and $B$ is an independent set.  In this case, $\mS=\{A,B\}$ is called a {\bf split partition}.
\end{definition}
Split graphs are a subset of the class of perfect graphs \cite{Roussel:2009jf,Chudnovsky06thestrong} and a superset of the threshold graphs \cite{Mahadev:1995uv}.  One of the striking things regarding split graphs is that they can be recognized solely by their degree sequences; a {\it degree sequence} $d$ is an integer sequence which is {\it graphic}, that is there is a labeled graph $G$ such that $\deg(v_{i}) = d_{i}$ for all $i=1,\ldots,|V(G)|$.  We will call the degree sequence of a split graph a split degree sequence.  The split degree sequence characterization states that a non-increasing degree sequence is the degree sequence of a split graph if and only if a particular Erd\H{o}s-Gallai inequality is satisfied with equality, where the Erd\H{o}s-Gallai inequalities give the necessary and sufficient conditions for an integer sequence to be graphic \cite{Erdos:1961ww}.  In \cite{Merris:2003vj} it is shown that degree sequences of split graphs are near the top of the partially ordered set of graphic integer sequences, with threshold graphs at the boundary of graphicality.

The fact that the degree sequence characterization of split graphs is intimately related to the concept of graphicality illustrates both their importance as well as their extendability to the directed case.  In particular, the Fulkerson inequalities are the directed version of the Erd\H{o}s-Gallai inequalities, that is, the Fulkerson inequalities give necessary and sufficient conditions for an integer-pair sequence to be digraphic \cite{Fulkerson:1960tv}.  In this paper, we define an extended version of split graphs, called {\it split digraphs}, that is a natural extension of both the structural characterization in Definition~\ref{def:splitg} as well as the degree sequence characterization using the Fulkerson inequalities.  

\subsection{Structure of split digraphs}

We first define a {\it split partition} of a directed graph $\DG$.  For this definition and discussion, given a digraph $\DG$ and vertex sets $X,Y\subset V(\DG)$, we define the subgraph $\DG[X,Y] = (X\cup Y, \DE[X,Y])$, where $\DE[X,Y] = \{(x,y)\in \DE(\DG) : x \in X \ \mbox{and} \ y \in Y\}$ and $(x,y)\in \DE(\DG)$ denotes an arc from $x$ to $y$.  When $X=Y$, we have the usual definition of an induced subgraph and will denote this by $\DG[X]$.
\begin{definition}[Split partition]
Given a digraph $\DG$, a vertex partition $\mS = \{\Spm,\Sp,\Sm,\S0\}$ (with possible empty sets) is called a {\bf split partition} of $\DG$ if and only if 

$(i)$ $\Spm$ is a clique and $\S0$ is an independent set,

$(ii)$ $\DG[\Sp]$ and $\DG[\Sm]$ are arbitrary subgraphs,

$(iii)$ there are all possible arcs from $\Sp$ to $\Spm\cup\Sm$ and from $\Spm$ to $\Sm$, and

$(iv)$ there are no arcs from $\Sm$ to $\Sp\cup\S0$ or from $\S0$ to $\Sp$.

\end{definition}
We can see immediately from $(i)$ alone that this definition includes split graphs as a special case.  Since vertex sets in $\mS$ can be empty, every digraph has the trivial split partitions $\mS=\{\Sp\}$ and $\mS=\{\Sm\}$.  This leads to the following definition of split digraphs.
\begin{definition}[Split digraph]
\label{def:splitdg}
A digraph $\DG$ is a {\bf split digraph} if and only if it has a nontrivial split partition.  
\end{definition}

\begin{figure}[t]
\centering
\includegraphics[width=4in]{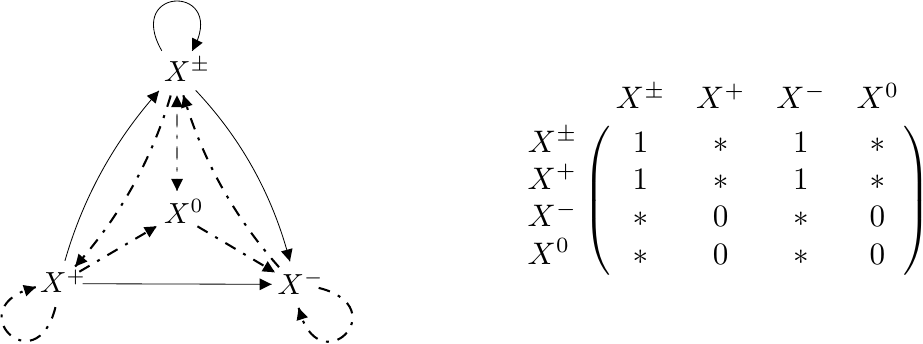}
\caption{{\bf Left:} Diagram showing the relations within and between the 4 possible vertex sets of a split partition $\mS = \{\Spm,\Sp,\Sm,\S0\}$.  Solid and dashed-dotted arrows denote all possible arcs and no constraints on arcs, respectively, while the absence of an arrow denotes no {\de}s.  $\Spm$ is a clique, $\S0$ an independent set, while $\DG[\Sp]$ and $\DG[\Sm]$ are arbitrary subgraphs. {\bf Right:} Adjacency matrix for $\DG$ in block form specified by the split partition $\mS$.}
\label{fig:mpart}
\end{figure}

The diagram in the left of Figure~\ref{fig:mpart} shows the relations within and between the four different classes of a split partition, with solid and dashed-dotted arrows denoting all possible arcs and no constraints on arcs, respectively, and the absence of an arrow denoting no arcs.  Another useful formulation of a digraph with a split partition is via its adjacency matrix, which is denoted in the right of Figure~\ref{fig:mpart} using the notation of $M$-partitions \cite{Feder:2003vp}.  In general, an $M$-partition of a digraph $\DG$ is a partition of the vertex-set $V(\DG)$ into $k$ disjoint sets $\{X_{1},\ldots,X_{k}\}$, where the {\de } constraints within and between sets are given by a $k\times k$ matrix $M$ with elements in $\{0,1,*\}$.  $M_{ii}$ equals $0$ or $1$ when $X_{i}$ is an independent set or clique, respectively, and is set to $*$ when $\DG[X_{i}]$ is an arbitrary subgraph.  Similarly, for $i\neq j$, $M_{ij}$ equal to $0$, $1$, or $*$ corresponds to $\DG[X_{i},X_{j}]$ having no {\de}s from $X_{i}$ to $X_{j}$, all {\de}s from $X_{i}$ to $X_{j}$, and no constraints on {\de}s from $X_{i}$ to $X_{j}$, respectively.

Similar to the undirected case, we will call the degree sequence of a split digraph a split degree sequence.  This paper shows that split digraphs have degree sequence characterizations analogous to the characterizations for split graphs \cite{Hammer:1981kn}, with the surprising result that a degree sequence for a directed graph is split if and only if {\it any} of the Fulkerson inequalities is satisfied with equality.  This is in direct contrast to the undirected case where a particular Erd\H{o}s-Gallai inequality has to be tight in order to be sufficient for a degree sequence to be split.

The structure of split digraphs has appeared in previous publications.  For example, their structure appears not only in the Fulkerson inequalities which show {\it when} an integer-pair sequence is graphic, but also in algorithms that determine {\it how} to construct a realization if the answer is affirmative.  More specifically, split digraphs appear in \cite{LaMar:2010tu} in a modification of the algorithm of Kleitman and Wang \cite{Kleitman:1973tm,Erdos:2010wi} to construct a directed graph realization from a degree sequence.  The algorithm of Kleitman and Wang is a directed analogue of the constructive Havel-Hakimi Algorithm and has appeared to have gone largely unnoticed in the literature.  Interestingly, split digraphs also give important structural information when using Markov chain Monte Carlo arc-switching techniques to sample directed graph realizations in an approximately uniform manner from a fixed degree sequence \cite{LaMar:2009tw,Berger:2010uv}.  In both of these cases, that is the extension of the algorithm of Kleitman and Wang and the sampling algorithm, a digraph composition operator is defined using split partitions that bears a striking resemblance to the composition operator used by Tyshkevich \cite{Tyshkevich:2000fh} to define the canonical split decomposition of undirected graphs.  

All of this shows a promise for split digraphs leading to many of the results enjoyed by split graphs, such as: a canonical split digraph decomposition, split degree sequences at the top of a suitably ordered set of digraphic degree sequences, and the extension of threshold graphs to the directed case.


\subsection{Undirected splittance and graphicality}

In this section, we introduce the two degree sequence characterizations for split graphs.  We spend the rest of the paper in Section~\ref{sec:digdeg} proving analogous versions of these degree sequence characterizations for split digraphs.  Hammer and Simeone \cite{Hammer:1981kn} determine two degree sequence characterizations of split graphs: one through the {\it splittance} of a graph and the other through equality of one of the Erd\H{o}s-Gallai inequalities, which give necessary and sufficient conditions for when an integer sequence is graphic.  We begin by discussing the concept of splittance, whose definition is given as follows.
\begin{definition}[Graph splittance \cite{Hammer:1981kn}]
\label{def:gsplit}
Define the {\bf splittance} $\sigma(G)$ of $G$ to be the minimum number of edges to add to or remove from $G$ in order to obtain a split graph.
\end{definition}
Clearly a graph $G$ is split if and only if $\sigma(G) = 0$.  The interesting fact is that the splittance $\sigma(G)$ can be written solely in terms of the degree sequence $d$ of $G$.  To see this, let the degree sequence $d$ be in non-increasing order, that is $d_{1} \geq \ldots \geq d_{N}$.  If we define the corrected Durfee number $m$ by
\[
m = m(d) = \max\{k : d_{k} \geq k-1\},
\]
and the splittance sequence $\{\sigma_{k}(d)\}_{k=0}^{N}$ by
\begin{equation}
\label{eq:splitseq}
\sigma_{k}(d) = \frac{1}{2}\left\{k(k-1) - \sum_{i=1}^{k}d_{i} + \sum_{i=k+1}^{N} d_{i}\right\},
\end{equation}
then we have the following theorem.
\begin{theorem}[Hammer and Simeone \cite{Hammer:1981kn}]
\label{thm:split1}
\[
\sigma(G) = \min_{k}\sigma_{k}(d) = \sigma_{m}(d).
\]
\end{theorem}
An important property of the splittance sequence \eqref{eq:splitseq} that leads to the second equality in Theorem~\ref{thm:split1} is that $\sigma_{k}(d)$ is non-increasing for $1 \leq k \leq m$ and strictly increasing for $m < k \leq N$.  Thus, $\sigma_{k}(d)$ has only one region of minima, which includes the corrected Durfee number $m$.  The following corollary of Theorem~\ref{thm:split1} is immediate and gives us the first degree sequence characterization of split graphs.
\begin{cor}[Hammer and Simeone \cite{Hammer:1981kn}]
\label{cor:split1}
If $d$ is a non-increasing degree sequence with $m$ its corrected Durfee number, then $d$ is split if and only if $\sigma_{m}(d) = 0$.
\end{cor}
The sufficiency of Corollary~\ref{cor:split1} can be illustrated as follows.  Suppose $G$ is split with non-increasing degree sequence $d$ and let $m$ be the corrected Durfee number for $d$.  We have $\mS = \{A,B\}$ defining a split partition of $G$, with the clique $A=\{x_{1},\ldots,x_{m}\}$ and the independent set $B=\{x_{m+1},\ldots,x_{N}\}$.  Clearly, all edges from $A$ either go back into $A$ or connect to $B$.  But the number of edges from $A$ is $\sum_{i=1}^{m}d_{i}$, while the number of edges going back into $A$ is $m(m-1)$ since $A$ is a clique, and the number of edges connecting to $B$ is $\sum_{i=m+1}^{N}d_{i}$ since $B$ is an independent set.  Thus, by \eqref{eq:splitseq}, $\sigma_{m}(d) = 0$.  In general, the choice of $\mS$ may not be unique.  For example, if there is a vertex in $A$ that has no connections with $B$, then $d_{m} = m-1$ and we also have the split partition $S^{\prime} = (A^{\prime},B^{\prime})$ with $A^{\prime} = \{x_{1},\ldots,x_{m-1}\}$ and $B^{\prime} = \{x_{m},\ldots,x_{N}\}$.
 
Hammer and Simeone also highlight a very interesting relationship between the splittance and the Erd\H{o}s-Gallai inequalities.
\begin{theorem}[Erd\H{o}s and Gallai \cite{Erdos:1961ww}]
\label{thm:erdos}
With $d$ a non-increasing integer sequence, $d$ is graphic if and only if $\sum_{i=1}^{N} d_{i}$ is even and for $k=1,\ldots,N$,
\[k(k-1) + \sum_{i=k+1}^{N} \min\{d_{i},k\} \geq \sum_{i=1}^{k} d_{i}.\]
\end{theorem}
For a non-increasing integer sequence, if we define the {\bf slack sequence} $\{s_{k}(d)\}_{k=0}^{N}$ by
\[s_{k}(d) = k(k-1) - \sum_{i=1}^{k} d_{i} + \sum_{i=k+1}^{N} \min\{d_{i},k\},\]  then Theorem~\ref{thm:erdos} can be restated as saying $d$ is graphic if and only if the slack sequence $s_{k}(d)$ is non-negative.  For any non-increasing integer sequence $d$ with $m$ its corrected Durfee number, by definition of $m$ we have $\min\{d_{i},m\} = d_{i}$ for $i=m+1,\ldots,N$.  Thus $2\sigma_{m}(d) = s_{m}(d)$, which by Corollary~\ref{cor:split1} gives us the following second degree sequence characterization of split graphs.
\begin{theorem}[Hammer and Simeone \cite{Hammer:1981kn}]
\label{thm:split2}
If $d$ is a non-increasing degree sequence with $m$ its corrected Durfee number, then $d$ is split if and only if $s_{m}(d) = 0$.
\end{theorem}
Thus, for a degree sequence to be split, the $m$-th Erd\H{o}s-Gallai inequality must be satisfied with equality.  This shows split sequences are somehow close to the boundary of graphicality.  

The remainder of the paper in Section~\ref{sec:digdeg} generalizes Corollary~\ref{cor:split1} and Theorem~\ref{thm:split2} to split digraphs and is outlined as follows.  Section~\ref{sec:digsplit} defines the splittance for a directed graph $\DG$ as the minimum number of arcs to add to or remove from $\DG$ to make $\DG$ a split digraph.  The digraph splittance can also be written solely in terms of the degree sequence of $\DG$, where a {\bf degree sequence} $d=\{d_{i}\}_{i=1}^{N} = \{(d_{i}^{+},d_{i}^{-})\}_{i=1}^{N}$ for a directed graph $\DG$ is an integer-pair sequence such that the out-degree and in-degree of vertex $v_{i}$ is $d^{+}_{i}$ and $d^{-}_{i}$, respectively.   We do this by defining a {\it splittance matrix} in Definition~\ref{def:splitmat}, similar to the splittance sequence in \eqref{eq:splitseq}.  We then prove in Theorem~\ref{thm:displit_split} an analogous version of Theorem~\ref{thm:split1} which states that the splittance of a directed graph is equal to the minimum of the splittance matrix (excluding entries corresponding to trivial split partitions).  This leads to the first degree sequence characterization of split digraphs in Corollary~\ref{cor:dschar1} that $\DG$ is split if and only if this minimum is zero.  Since the splittance matrix may have multiple disconnected regions of minima, in contrast to the splittance sequence \cite{Hammer:1981kn}, the corrected Durfee number $m$ does not have a direct extension to degree sequences for directed graphs.  However, in Definition~\ref{def:maxseq} we define an extension which we call {\it maximal sequences} that do share similar properties to the corrected Durfee number, and which exhibit interesting structure relative to the splittance matrix, as can be seen in Corollary~\ref{cor:minsig} and Lemma~\ref{lem:slacksplit} in Section~\ref{sec:digslack}.  This allows us to prove the second degree sequence characterization of split digraphs in Corollary~\ref{cor:dschar2}, which is analogous to Theorem~\ref{thm:split2}, stating that a digraph $\DG$ is split if and only if its degree sequence satisfies {\it any} of the Fulkerson inequalities with equality, where the Fulkerson inequalities give necessary and sufficient conditions for an integer-pair sequence to be digraphic \cite{Fulkerson:1960tv}.  This is stronger than the undirected case, which states that an undirected graph is split if and only if the $m$-th Erd\H{o}s-Gallai inequality is satisfied with equality, where $m$ is the corrected Durfee number.

\section{Degree sequence characterizations of split digraphs}
\label{sec:digdeg}

In working with the degree sequence characterizations for split graphs using both the splittance as well as the slack sequences, the degree sequence $d$ needed to be non-increasing.  In the directed case, we need degree sequences to be non-increasing as well, in particular under the lexicographical ordering of the first or second coordinate.  Given an integer-pair sequence $d = \{(d^{+}_{i}, d^{-}_{i})\}_{i=1}^{N}$, the permutation sequence $a=\{a_{i}\}_{i=1}^{N}$ will be called a {\bf positive lexicographical ordering} if and only if $d^{+}_{a_{i}} \geq d^{+}_{a_{i+1}}$, with $d^{-}_{a_{i}} \geq d^{-}_{a_{i+1}}$ when $d^{+}_{a_{i}} = d^{+}_{a_{i+1}}$.  Similarly, we call the permutation sequence $b=\{b_{i}\}_{i=1}^{N}$ a {\bf negative lexicographical ordering} if and only if $d^{-}_{b_{i}} \geq d^{-}_{b_{i+1}}$, with $d^{+}_{b_{i}} \geq d^{+}_{b_{i+1}}$ when $d^{-}_{b_{i}} = d^{-}_{b_{i+1}}$.

We need one more subtle property of our ordering which states that if $d_{i}=d_{j}$, then $d_{i}$ is before $d_{j}$ in the positive lexicographical ordering if and only if $d_{i}$ is before $d_{j}$ in the negative lexicographical ordering.
\begin{definition}[Proper ordering]
An integer-pair sequence $d$ is said to be {\bf properly ordered} with positive and negative lexicographical orderings $a$ and $b$ such that if $d_{i} = d_{j}$, where $i=a_{m}=b_{n}$ and $j=a_{m^{\prime}}=b_{n^{\prime}}$, then $m < m^{\prime}$ if and only if $n < n^{\prime}$.
\end{definition}

\subsection{Splittance of a directed graph}
\label{sec:digsplit}

We begin this section with a generalization of graph splittance in Definition~\ref{def:gsplit} to digraphs.
\begin{definition}[Digraph splittance]
\label{def:dgsplit}
Define $\sigma(\vec{G})$ to be the minimum number of arcs to be added to or removed from $\vec{G}$ in order to obtain a split digraph.
\end{definition}
The following defines two measures on arbitrary partitions $\mS=\{\Spm,\Sp,\Sm,\S0\}$, which as we show in Lemma~\ref{lem:minsplitpart} tell us how close $\mS$ is to being a split partition.
\begin{definition}[Split partition measures]
\label{def:splitmeas}
Let $\mS = \{\Spm,\Sp,\Sm,\S0\}$ be an arbitrary vertex partition. With $k\equiv |\Spm|+|\Sp|$ and $l\equiv |\Spm|+|\Sm|$, define the {\it split partition measures} $\bar{\sm}(\mS)$ and $\ubar{\sm}(\mS)$ as
\[
\begin{split}
\bar{\sm}(\mS) & = |\Spm|\bigl(k-1\bigr) + |\Sm|k + \sum_{\Sp\cup\S0}d_{x}^{-} - \sum_{\Spm\cup\Sp}d_{x}^{+},\\
\ubar{\sm}(\mS) & = |\Spm|\bigl(l-1\bigr) + |\Sp|l + \sum_{\Sm\cup\S0}d_{x}^{+} - \sum_{\Spm\cup\Sm}d_{x}^{-}.
\end{split}
\]
\end{definition}
By the definition of $k$ and $l$ in Definition~\ref{def:splitmeas}, we have the equivalent formulation
\begin{equation}
\label{eq:splitmeas2}
\begin{split}
\bar{\sm}(\mS) & = kl - |\Spm| + \sum_{\Sp\cup\S0}d^{-}_{x} - \sum_{\Spm\cup\Sp}d^{+}_{x},\\
\ubar{\sm}(\mS) & = kl - |\Spm| + \sum_{\Sm\cup\S0}d^{+}_{x} - \sum_{\Spm\cup\Sm}d^{-}_{x}.
\end{split}
\end{equation}
\begin{lem}
\label{lem:smequiv}
For an arbitrary partition $\mS = \{\Spm,\Sp,\Sm,\S0\}$, we have
\[
\bar{\sm}(\mS) = \ubar{\sm}(\mS).
\]
\end{lem}

\bnprf
Let $\mS=\{\Spm,\Sp,\Sm,\S0\}$ be a partition with $k=|\Spm|+|\Sp|$ and $l=|\Spm|+|\Sm|$.  Using \eqref{eq:splitmeas2}, we have
\begin{eqnarray*}
\bar{\sm}(\mS) - \ubar{\sm}(\mS) & = & kl - |\Spm| + \sum_{\Sp\cup\S0}d_{x}^{-} - \sum_{\Spm\cup\Sp}d_{x}^{+} \\
                                            &   & - \left\lbrack kl - |\Spm| + \sum_{\Sm\cup\S0}d_{x}^{+} - \sum_{\Spm\cup\Sm}d_{x}^{-}\right\rbrack \\
                                            & = & \sum_{V}d_{x}^{-} - \sum_{V}d_{x}^{+} \\
                                            & = & 0
\end{eqnarray*}
\edprf
We thus speak of the split partition measure $\sm(\mS)$ and work with $\sm \equiv \bar{\sm}$ in all proofs that follow.  To see that $\sm(\mS) = 0$ for a split partition $\mS$, consider Figure~\ref{fig:mpart} which shows that all arcs from $\Spm\cup\Sp$ include all possible arcs from $\Spm\cup\Sp$ into $\Spm\cup\Sm$, as well as the arcs from $\Spm\cup\Sp$ into $\Sp\cup\S0$.  However, the total number of arcs from $\Spm\cup\Sp$ is given by $\sum_{\Spm\cup\Sp}d_{x}^{+}$, while the total number of all possible arcs from $\Spm\cup\Sp$ into $\Spm\cup\Sm$ is given by $|\Spm|(k-1)+|\Sm|k$, and the total number of arcs from $\Spm\cup\Sp$ into $\Sp\cup\S0$ is given by $\sum_{\Sp\cup\S0}d_{x}^{-}$.  This gives $\sm(\mS) = 0$ by Definition~\ref{def:splitmeas}.  

The next lemma shows in fact that for an arbitrary partition $\mS=\{\Spm,\Sp,\Sm,\S0\}$, $\sigma(\mS)$ gives the minimal number of arcs to add to or remove from $\DG$ in order for $\mS$ to be a split partition.  In particular, this implies $\sigma(\DG) = \min_{\mS}\sm(\mS)$ when minimizing over nontrivial partitions $\mS$, that is, partitions such that $\mS\neq\{\Sp\}$ or $\{\Sm\}$.  Thus, we have
\begin{lem}
For a partition $\mS=\{\Spm,\Sp,\Sm,\S0\}$, $\sm(\mS)$ gives the minimal number of arcs to add to or remove from $\DG$ in order for $\mS$ to be a split partition.
\label{lem:minsplitpart}
\end{lem}

\bnprf
Given a digraph $\DG$, recall the notation $\DE[X,Y] = \{(x,y)\in \DE(\DG) : x \in X \ \mbox{and} \ y \in Y\}$, where $X,Y\subset V(\DG)$ and $(x,y)\in \DE(\DG)$ denotes an arc from $x$ to $y$.  For a partition $\mS = \{\Spm,\Sp,\Sm,\S0\}$ with $k=|\Spm|+|\Sp|$ and $l=|\Spm|+|\Sm|$, it is easily seen that
\begin{eqnarray*}
\sum_{\Sp\cup\S0}d^{-}_{x} & = & \Bigl|A\bigl[\Sm\cup\S0,\Sp\cup\S0\bigr]\Bigr| + \Bigl|A\bigl[\Spm\cup\Sp,\Sp\cup\S0\bigr]\Bigr| \\
\sum_{\Spm\cup\Sp}d^{+}_{x} & = & \Bigl|A\bigl[\Spm\cup\Sp,\Spm\cup\Sm\bigr]\Bigr| + \Bigl|A\bigl[\Spm\cup\Sp,\Sp\cup\S0\bigr]\Bigr|,
\end{eqnarray*}
and thus
\[
\sum_{\Sp\cup\S0}d^{-}_{x} - \sum_{\Spm\cup\Sp}d^{+}_{x} = \Bigl|A\bigl[\Sm\cup\S0,\Sp\cup\S0\bigr]\Bigr| - \Bigl|\bigl[\Spm\cup\Sp,\Spm\cup\Sm\bigr]\Bigr|.
\]
This leads to
\[
\sm(\mS) = |\Spm|\bigl(k-1\bigr) + |\Sm|k - \Bigl|A\bigl[\Spm\cup\Sp,\Spm\cup\Sm\bigr]\Bigr| + \Bigl|A\bigl[\Sm\cup\S0,\Sp\cup\S0\bigr]\Bigr|.
\]
Note that the first three terms give the number of arcs to add to $\DG$ for there to be all arcs from $\Spm\cup\Sp$ to $\Spm\cup\Sm$, while the last term gives the number of arcs to remove from $\DG$ so that there are no arcs from $\Sm\cup\S0$ to $\Sp\cup\S0$.  The resulting digraph after addition and removal of these arcs will have $\mS$ as a split partition.  By the definition of a split partition, this is the minimal number of arcs since any other arc that is added to or removed from $\DG$ is unnecessary, as illustrated by the asterisks in the adjacency matrix in Figure~\ref{fig:mpart}.
\edprf
\begin{cor}
\label{cor:splitpart}
A partition $\mS=\{\Spm,\Sp,\Sm,\S0\}$ is a split partition if and only if $\sm(\mS) = 0$.
\end{cor}
It helps at this point to make more explicit the subtle distinction between $\sigma(\mS)$, which tells how far $\mS$ is from being a split partition, and $\sigma(\DG)$, which gives the distance to the closest split digraph.  More specifically, since split digraphs are defined in terms of nontrivial split partitions, we will be minimizing $\sigma(\mS)$ over all nontrivial partitions with $\mS\neq\{\Sp\}$ and $\mS\neq\{\Sm\}$.  Thus, from Lemma~\ref{lem:minsplitpart}, we have the following corollary.
\begin{cor}
\label{cor:minsplit1}
Minimizing over nontrivial partitions $\mS=\{\Spm,\Sp,\Sm,\S0\}$, we have
\[
\sigma(\vec{G}) = \min_{\mS}\sm(\mS).
\]
\end{cor}
In Theorem~\ref{thm:displit_split}, we show for each fixed index pair $(k,l)\in[0,N]\times[0,N]$, there is a partition $\mX_{kl}$ such that 
\begin{equation}
\label{eq:indmin}
\sigma(\mX_{kl}) = \min_{\stackrel{|\Spm|+|\Sp|=k}{\scriptscriptstyle |\Spm|+|\Sm|=l}}\sm(\mS).
\end{equation}
Note that the trivial partitions have $\Spm=\emptyset$, and thus $\mS=\{\Sp\}$ and $\mS=\{\Sm\}$ correspond to the index-pairs $(k,l)=(N,0)$ and $(k,l)=(0,N)$, respectively.  Putting this all together, we have
\begin{equation}\label{eq:steps}
\begin{split}
\sigma(\DG) & = \min_{\mS}\sm(\mS), \quad \mS \ \mbox{nontrivial} \\
            & = \min_{(k,l)\notin\{(0,N),(N,0)\}}\min_{\stackrel{|\Spm|+|\Sp|=k}{\scriptscriptstyle |\Spm|+|\Sm|=l}}\sm(\mS) \\
            & = \min_{(k,l)\notin\{(0,N),(N,0)\}}\sm(\mX_{kl}) \\
            & = \min_{(k,l)\notin\{(0,N),(N,0)\}}\Sig_{kl},
\end{split}
\end{equation}
where $\Sig_{kl}\equiv\sm(\mX_{kl})$ is called the {\it splittance matrix} and is the directed extension of the splittance sequence in \eqref{eq:splitseq}.  These special partitions $\mX_{kl}$ are called {\it induced partitions}, which are defined along with a formal definition of the splittance matrix in the following way.
\begin{definition}[Induced partitions]
\label{def:indpart}
Suppose $d$ is properly ordered with $a$ and $b$ the corresponding positive and negative lexicographical orderings, respectively.  For $(k,l)\in [0,N] \times [0,N]$, let $\mA_{k} = \{a_{i}\}_{i=1}^{k}$ and $\mB_{l} = \{b_{i}\}_{i=1}^{l}$.  The partition induced by $(k,l)$ is given by $\mX_{kl} = \{\Xpm,\Xp,\Xm,\X0\}$ such that
\begin{eqnarray*}
\Xpm & = & V_{\mA_{k}\cap\mB_{l}}, \\
\Xp  & = & V_{\mA_{k}\setminus\mB_{l}}, \\
\Xm  & = & V_{\mB_{l}\setminus\mA_{k}}, \\
\X0  & = & V_{\{1,\ldots,N\}\setminus(\mA_{k}\cup\mB_{l})},
\end{eqnarray*}
where the notation $V_{\mathcal{K}}$ denotes the set of all labeled vertices indexed by the set $\mathcal{K}$.
\end{definition}
\begin{definition}[Splittance matrix]
\label{def:splitmat}
Suppose $d$ is properly ordered.  For each $(k,l)\in [0,N] \times [0,N]$ and corresponding induced partition $\mX_{kl}$, we define the splittance matrix $\Sig\equiv\Sig(d)$ such that $\Sig_{kl} = \sm(\mX_{kl})$.
\end{definition}
It is important to note that the splittance matrix has starting indices at 0 and not 1.  

In the chain of equalities in \eqref{eq:steps}, for the equality
\[\min_{\stackrel{|\Spm|+|\Sp|=k}{\scriptscriptstyle |\Spm|+|\Sm|=l}}\sm(\mS)=\sm(\mX_{kl})\]
to be satisfied, it is critical that $d$ be properly ordered in constructing the induced partitions $\mX_{kl}$.  For example, for the degree sequence 
\[
d = \left(
\begin{array}{ccc}
d_{1}^{+} & \cdots & d_{4}^{+} \\
d_{1}^{-} & \cdots & d_{4}^{-}
\end{array}
\right) = \left(
\begin{array}{cccc}
1 & 1 & 2 & 2 \\
2 & 2 & 1 & 1 
\end{array}
\right),
\]
if we choose the positive lexicographical ordering $a=(3 \ 4 \ 1 \ 2)$ and negative lexicographical ordering $b=(2 \ 1 \ 3 \ 4)$, which does {\it not} give a proper ordering, then the partition $\mS$ ``induced'' by $(k,l)=(3,1)$ (which is $\Spm=\emptyset$, $\Sp=\{1,3,4\}$, $\Sm=\{2\}$ and $\S0=\emptyset$) leads to $\sigma(\mS) = 2$.  However, when we use the proper ordering with $a=(3 \ 4 \ 1 \ 2)$ and $b=(1 \ 2 \ 3 \ 4)$, then the induced partition $\mX_{31}$ gives $\sigma(\mX_{31}) = 1$.

The main theorem of this section is Theorem~\ref{thm:displit_split} showing
\begin{equation}
\label{eq:displit_split}
\sigma(\vec{G}) = \min_{(k,l)\notin\{(N,0),(0,N)\}}\Sig_{kl},
\end{equation}
which by \eqref{eq:steps} reduces to showing \eqref{eq:indmin}.  Note again that the index-pairs $(0,N)$ and $(N,0)$ correspond to trivial split partitions and therefore $\Sig_{0N} = \Sig_{N0} = 0$.  Thus, these corners of $\Sig$ are not used in computing the splittance, as seen in \eqref{eq:displit_split}.

Before proceeding to Theorem~\ref{thm:displit_split}, it is instructive to give some examples.  Consider the degree sequence 
\begin{equation}
\label{eq:ex1}
d = \left(
\begin{array}{ccccc}
2 & 3 & 4 & 1 & 0 \\
1 & 2 & 2 & 2 & 3
\end{array}
\right),
\end{equation}
which as we will see below is split.  A particular labeled realization $\DG$ of $d$ is given by
\smallskip
\begin{center}
\includegraphics{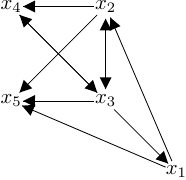}
\end{center}
\smallskip
For a proper ordering of $d$, define the positive lexicographical ordering $a = (3\,2\,1\,4\,5)$ and negative lexicographical ordering $b = (5\,3\,2\,4\,1)$.  The splittance matrix $\Sig\equiv\Sig(d)$ is given by
\[
\Sig = \left(\begin{array}{cccccc}
10 & 7  & 5  & 3  & 1  & 0 \cr
6  & 4  & 2  & 1  & 0  & 0 \cr
3  & 2  & 1  & 0  & 0  & 1 \cr
1  & 1  & 1  & 1  & 2  & 3 \cr
0  & 1  & 2  & 3  & 4  & 6 \cr
0  & 1  & 3  & 5  & 7  & 10
\end{array}\right).
\]
There are 5 nontrivial zeros in $\Sig$ corresponding to 5 different induced split partitions.  For example, for the index pair $(k,l) = (2,3)$, which corresponds to the third row and fourth column of $S$ since the indices start at 0, we can define the sets $\mathcal{A}_{2} = \{a_{1},a_{2}\} = \{3, 2\}$ and $\mathcal{B}_{3} = \{b_{1},b_{2},b_{3}\} = \{5, 3, 2\}$.  This index pair induces a split partition $\mX_{23}$ as follows
\begin{eqnarray*}
\Xpm & = & V_{\mA_{2}\cap \mB_{3}} = \{x_{2},x_{3}\}, \\
\Xp  & = & V_{\mA_{2} \setminus \mB_{3}} = \emptyset, \\
\Xm  & = & V_{\mB_{3} \setminus \mA_{2}} = \{x_{5}\}, \\
\X0  & = & V_{\{1,\ldots,5\}\setminus(\mA_{2}\cup\mB_{3})} = \{x_{1},x_{4}\}.
\end{eqnarray*}
As mentioned previously, the splittance sequence \eqref{eq:splitseq} has one region of minima, which includes the corrected Durfee number.  This example illustrates, however, that there can be multiple regions of minima in $\Sig$ separated by nonzero splittance.

To illustrate the connections between undirected and directed splittance, consider the degree sequence $d = (4\ 3\ 3\ 3\ 3)$, with the directed extension of $d$ given by
\begin{equation}
\label{eq:dirext}
\left(\begin{array}{ccccc} 4 & 3 & 3 & 3 & 3 \\ 4 & 3 & 3 & 3 & 3 \end{array}\right).
\end{equation}
The splittance sequence for $d$ is $\sigma(d) = (8\ 4\ 2\ 1\ 1\ 2)$, while the splittance matrix for \eqref{eq:dirext} is given by
\[
\Sig = \left(\begin{array}{cccccc}
16 & 12  & 9  & 6  & 3  & 0 \cr
12  & 8  & 6  & 4  & 2  & 0 \cr
9  & 6  & 4  & 3  & 2  & 1 \cr
6  & 4  & 3  & 2  & 2  & 2 \cr
3  & 2  & 2  & 2  & 2  & 3 \cr
0  & 0  & 1  & 2  & 3  & 4
\end{array}\right).
\]
In this case, $\Sig$ is symmetric and $\sigma(d) = \frac{1}{2}\mbox{diag}(\Sig)$, with the factor of $\frac{1}{2}$ accounting for the identification of two arcs for every edge.  This is an interesting example in that by Theorem~\ref{thm:split1}, $d$ is not split since $\sigma(d)$ is nonzero everywhere, while its directed extension \eqref{eq:dirext} {\it is} split since $S_{51} = S_{15} = 0$.

We see from Definition~\ref{def:indpart} that $V_{\mA_{k}} = \Xpm\cup\Xp$ and $V_{\mB_{l}} = \Xpm\cup\Xm$, which gives: 
\begin{subequations}
\begin{align}
& \bullet \ x\in\Xpm\cup\Xp \ \mbox{and} \ y\in\Xm\cup\X0 \Longrightarrow d_{x} \geq d_{y} \ \mbox{in the positive lexicographical ordering}, \label{eq:Pineq} \\
& \bullet \ x\in\Xpm\cup\Xm \ \mbox{and} \ y\in\Xp\cup\X0 \Longrightarrow d_{x} \geq d_{y} \ \mbox{in the negative lexicographical ordering}. \label{eq:Nineq}
\end{align}
\end{subequations}
To prove \eqref{eq:displit_split} in Theorem~\ref{thm:displit_split}, we need the following inequalities of induced partitions, which rely heavily on the proper ordering of $d$ and give conditions on when strict inequality in \eqref{eq:Pineq} and \eqref{eq:Nineq} occurs.

\begin{lem}
\label{lem:ineq}
Suppose $d$ is properly ordered and $\mX = \{\Xpm, \Xp, \Xm, \X0\}$ is an induced partition.  For $x\in\Xp$, $y\in\Xm$, $z\in\Xpm$ and $w\in\X0$, we have
\begin{subequations}
\begin{align}
d_{x}^{+} > d_{y}^{+}, \label{eq:ineq1} \\
d_{y}^{-} > d_{x}^{-}, \label{eq:ineq2}
\end{align}
\begin{align}
d_{z}^{-} = d_{x}^{-} \Rightarrow d_{z}^{+} > d_{y}^{+}, \label{eq:ineq3} \\
d_{z}^{+} = d_{y}^{+} \Rightarrow d_{z}^{-} > d_{x}^{-}, \label{eq:ineq4}
\end{align}
\begin{align}
d_{y}^{-} = d_{w}^{-} \Rightarrow d_{x}^{+} > d_{w}^{+}, \label{eq:ineq5} \\
d_{x}^{+} = d_{w}^{+} \Rightarrow d_{y}^{-} > d_{w}^{-}. \label{eq:ineq6}
\end{align}
\end{subequations}
\end{lem}

\bnprf
Let $a$ and $b$ be the positive and negative lexicographical orderings, respectively, that give a proper ordering of $d$.  We now prove \eqref{eq:ineq1} and \eqref{eq:ineq3}, with the others following similarly.  For \eqref{eq:ineq1}, let $x\in\Xp$ and $y\in\Xm$.  By \eqref{eq:Pineq}, $d_{x}^{+} \geq d_{y}^{+}$.  If $d_{x}^{+} > d_{y}^{+}$, then we're done, so suppose $d_{x}^{+} = d_{y}^{+}$ with $d_{x}^{-} \geq d_{y}^{-}$.  But by \eqref{eq:Nineq}, $d_{y} \geq d_{x}$ in the negative lexicographical ordering, which implies $d_{y}^{-} \geq d_{x}^{-}$, and thus $d_{x}=d_{y}$.  Suppose $x=a_{m}=b_{n}$ and $y=a_{m^{\prime}}=b_{n^{\prime}}$.  By \eqref{eq:Pineq}, we have $m < m^{\prime}$.  But by the proper ordering of $d$ and the fact that $d_{x} = d_{y}$, we must have $n < n^{\prime}$.  However, by \eqref{eq:Nineq} and since $y \in \Xm$, $n < n^{\prime} \Rightarrow x \in \Xpm\cup\Xm$, which is a contradiction.

Now consider \eqref{eq:ineq3} and let $x\in\Xp$, $y\in\Xm$ and $z\in\Xpm$ such that $d_{z}^{-} = d_{x}^{-}$ with $d_{z}^{+} \leq d_{y}^{+}$.  Since $z\in\Xpm$, $x\in\Xp$ and $d_{z}^{-}=d_{x}^{-}$, by \eqref{eq:Nineq} we have $d_{z}^{+} \geq d_{x}^{+}$.  But this means $d_{x}^{+} \leq d_{y}^{+}$, which contradicts \eqref{eq:ineq1}.
\edprf

\begin{thm}
\[
\sigma(\vec{G}) = \min_{(k,l)\notin \{(0,N),(N,0)\}}\Sig_{kl}
\]
\label{thm:displit_split}
\end{thm}

\bnprf
By the series of equalities in \eqref{eq:steps}, we only need to show
\begin{equation}
\label{eq:minsplit}
\min_{\stackrel{|\Spm|+|\Sp|=k}{\scriptscriptstyle |\Spm|+|\Sm|=l}}\sm(\mS) = \Sig_{kl}.
\end{equation}
We will work with $\Sig_{kl} \equiv \bar{\sm}(\mX_{kl})$ and for simplicity drop the subscripts to $\mX$.  We thus want to show $\sm(\mS) \geq \sm(\mX)$ for all partitions $\mS=\{\Spm,\Sp,\Sm,\S0\}$ such that $|\Spm|+|\Sp| = k$ and $|\Spm|+|\Sm| = l$.  Letting $\mS$ be such a partition, then by \eqref{eq:Nineq} and \eqref{eq:Pineq} we have
\begin{equation}
\label{eq:indpartineq}
\sum_{\Sp\cup\S0}d^{-}_{x} \geq \sum_{\Xp\cup\X0}d^{-}_{x} \qquad \mbox{and} \qquad \sum_{\Spm\cup\Sp}d^{+}_{x} \leq \sum_{\Xpm\cup\Xp}d^{+}_{x}.
\end{equation}
If $|\Xpm| \geq |\Spm|$, then by \eqref{eq:splitmeas2} and \eqref{eq:indpartineq} we have $\sm(\mS) \geq \sm(\mX)$.  We need to do more work however when $|\Xpm| < |\Spm|$.  We have
\begin{subequations}
\begin{eqnarray}
\sm(\mS)-\sm(\mX) & = & kl-|\Spm| + \sum_{\Sp\cup\S0}d_{x}^{-} - \sum_{\Spm\cup\Sp}d_{x}^{+} - \left\lbrack kl-|\Xpm| + \sum_{\Xp\cup\X0}d_{x}^{-} - \sum_{\Xpm\cup\Xp}d_{x}^{+}\right\rbrack \nonumber \\
              & = & \left(\sum_{\Sp\cup\S0}d_{x}^{-} - \sum_{\Xp\cup\X0}d_{x}^{-}\right) + \left(\sum_{\Xpm\cup\Xp}d_{x}^{+} - \sum_{\Spm\cup\Sp}d_{x}^{+}\right) + |\Xpm|-|\Spm| \nonumber \\
              & = & \left(\sum_{(\Sp\cup\S0)\setminus(\Xp\cup\X0)}d_{x}^{-} - \sum_{(\Xp\cup\X0)\setminus(\Sp\cup\S0)}d_{x}^{-}\right) + \nonumber \\
              &   & \left(\sum_{(\Xpm\cup\Xp)\setminus(\Spm\cup\Sp)}d_{x}^{+} - \sum_{(\Spm\cup\Sp)\setminus(\Xpm\cup\Xp)}d_{x}^{+}\right) + |\Xpm|-|\Spm| \nonumber \\
              & = & \left(\sum_{(\Xpm\cup\Xm)\cap(\Sp\cup\S0)}d_{x}^{-} - \sum_{(\Xp\cup\X0)\cap(\Spm\cup\Sm)}d_{x}^{-}\right) + \label{eq:key1} \\
              &   & \left(\sum_{(\Xpm\cup\Xp)\cap(\Sm\cup\S0)}d_{x}^{+} - \sum_{(\Xm\cup\X0)\cap(\Spm\cup\Sp)}d_{x}^{+}\right) + |\Xpm|-|\Spm|. \label{eq:key2}
\end{eqnarray}
\end{subequations}
Note that the two sums in \eqref{eq:key2} have the same number of terms since
\begin{eqnarray*}
|(\Xpm\cup\Xp)\cap(\Sm\cup\S0)| & = & |\Xpm\cup\Xp| - |(\Xpm\cup\Xp)\cap (\Spm\cup\Sp)| \\
                                & = & |\Spm\cup\Sp| - |(\Xpm\cup\Xp)\cap(\Spm\cup\Sp)| \\
                                & = & |(\Xm\cup\X0)\cap(\Spm\cup\Sp)|.
\end{eqnarray*}
The same can be shown for the two sums in \eqref{eq:key1}.  Now let $n = |\Spm|-|\Xpm| > 0$ and 
\begin{align}
\Omega^{-} & = \left(\sum_{(\Xpm\cup\Xm)\cap(\Sp\cup\S0)}d_{x}^{-} - \sum_{(\Xp\cup\X0)\cap(\Spm\cup\Sm)}d_{x}^{-}\right), \nonumber \\
\Omega^{+} & = \left(\sum_{(\Xpm\cup\Xp)\cap(\Sm\cup\S0)}d_{x}^{+} - \sum_{(\Xm\cup\X0)\cap(\Spm\cup\Sp)}d_{x}^{+}\right), \nonumber
\end{align}
so that
\[\sm(\mS)-\sm(\mX) = \Omega^{-}+\Omega^{+}-n.\]
We need to show $\Omega^{-}+\Omega^{+} \geq n$.  Note that by \eqref{eq:Nineq} and \eqref{eq:Pineq}, $\Omega^{-} \geq 0$ and $\Omega^{+} \geq 0$ and thus $\Omega^{-} + \Omega^{+} \geq 0$.  
Since $|\Xpm|+|\Xp| = |\Spm|+|\Sp|$, we have $n = |\Spm|-|\Xpm| = |\Xp|-|\Sp|$, and therefore there are at least $n$ elements in the set
$\Xp\setminus\Sp$.  As $|\Xpm|+|\Xm| = |\Spm|+|\Sm|$, by the same argument there are at least $n$ elements in the set $\Xm\setminus\Sm$.  Let $\{x_{1},\ldots,x_{n}\} \subset \Xp\setminus\Sp$ and $\{y_{1},\ldots,y_{n}\}\subset \Xm\setminus\Sm$.  Our technique is to go through all the pairings $(x_{i},y_{i})$ and use the strict inequalities in \eqref{eq:ineq1}--\eqref{eq:ineq6} to show $\Omega^{-} + \Omega^{+} \geq n$.  
Using the fact that $\Xp\setminus\Sp = \Xp\cap(\Spm\cup\Sm\cup\S0)$ and $\Xm\setminus\Sm = \Xm\cap(\Spm\cup\Sp\cup\S0)$, we have the following four cases:
\begin{subequations}
\begin{align}
x_{i}\in\Xp\cap\Spm & \quad \ \mbox{and} \ \quad y_{i}\in\Xm\cap\Spm, \label{eq:case1} \\
x_{i}\in\Xp\cap\Spm & \quad \ \mbox{and} \ \quad y_{i}\in\Xm\cap(\Sp\cup\S0), \label{eq:case2} \\
x_{i}\in\Xp\cap(\Sm\cup\S0) & \quad \ \mbox{and} \ \quad y_{i}\in\Xm\cap\Spm, \label{eq:case3} \\
x_{i}\in\Xp\cap(\Sm\cup\S0) & \quad \ \mbox{and} \ \quad y_{i}\in\Xm\cap(\Sp\cup\S0). \label{eq:case4}
\end{align}
\end{subequations}
We break case \eqref{eq:case4} into 4 sub-cases as follows:
\begin{subequations}
\begin{align}
x_{i}\in\Xp\cap\S0 & \quad \ \mbox{and} \ \quad y_{i}\in\Xm\cap\S0, \label{eq:case4-1} \\
x_{i}\in\Xp\cap\Sm & \quad \ \mbox{and} \ \quad y_{i}\in\Xm\cap\S0, \label{eq:case4-2} \\
x_{i}\in\Xp\cap\S0 & \quad \ \mbox{and} \ \quad y_{i}\in\Xm\cap\Sp, \label{eq:case4-3} \\
x_{i}\in\Xp\cap\Sm & \quad \ \mbox{and} \ \quad y_{i}\in\Xm\cap\Sp. \label{eq:case4-4}
\end{align}
\end{subequations}
All cases that we must deal with include \eqref{eq:case1}--\eqref{eq:case3} and \eqref{eq:case4-1}--\eqref{eq:case4-4}.  Cases \eqref{eq:case1} and \eqref{eq:case4-1} must be dealt with differently, since in contrast to the other cases they give one term in $\Omega^{-}$ and another term in $\Omega^{+}$.
First, if there is an $(x_{i},y_{i})$ and $(x_{j},y_{j})$ satisfying \eqref{eq:case1} and \eqref{eq:case4-1}, respectively, then we can switch the pairings to have $(x_{i},y_{j})$ and $(x_{j},y_{i})$ satisfying \eqref{eq:case2} and \eqref{eq:case3}, respectively.  After such a re-pairing, we will have no pairings that satisfy \eqref{eq:case1} and/or no pairings that satisfy \eqref{eq:case4-1}.  Suppose there are pairings that satisfy \eqref{eq:case1} and none that satisfy \eqref{eq:case4-1} (the opposite case can be proved similarly), and let $(x,y)$ be such a pairing.  We will define a partition $\hat{\mS}$ such that $|\hSpm\cup\hSp| = k$ and $|\hSpm\cup\hSm| = l$ with $|\hSpm| = |\Spm|-1$.  We have
\[
\sm(\mS)-\sm(\mX) = \sm(\mS) - \sm(\hat{\mS}) + \sm(\hat{\mS})-\sm(\mX).
\]
We show below that the constructed partition $\hat{\mS}$ will satisfy $\sm(\mS) - \sm(\hat{\mS}) \geq 0$ and thus 
\[\sm(\mS)-\sm(\mX) \geq \sm(\hat{\mS})-\sm(\mX).\]
We can then apply this procedure recursively to remove all pairings satisfying \eqref{eq:case1}.  Supposing for the sake of argument that $\hat{\mS}$ is the final partition with no pairings in \eqref{eq:case1}, if we can then show $\sm(\hat{\mS})-\sm(\mX) \geq 0$, then we'll be done.

Since $|\Spm|+|\Sp| = |\Xpm|+|\Xp|$, by the definition of a partition we must have $|\Sm|+|\S0| = |\Xm|+|\X0|$.  This implies $|\S0|-|\X0| = |\Xm|-|\Sm| = n$, and thus there are at least $n$ elements in $\S0\setminus\X0$.  We have three more cases for an element $z \in \S0\setminus\X0=\S0\cap(\Xpm\cup\Xp\cup\Xm)$, with each case and the corresponding definition of $\hat{\mS}$ defined below:
\begin{subequations}
\begin{align}
z\in\Xp\cap\S0 & \ \quad \Rightarrow \ \quad \hat{\mS} = \{\Spm-y, \Sp+z, \Sm+y, \S0-z\}, \label{eq:Xp} \\
z\in\Xm\cap\S0 & \ \quad \Rightarrow \ \quad \hat{\mS} = \{\Spm-x, \Sp+x, \Sm+z, \S0-z\}, \label{eq:Xm} \\
z\in\Xpm\cap\S0 & \ \quad \Rightarrow \ \quad \hat{\mS} = \{\Spm-x-y+z, \Sp+x, \Sm+y, \S0-z\}. \label{eq:Xpm}
\end{align}
\end{subequations}
If in \eqref{eq:key1} and \eqref{eq:key2} we replace $\mX$ with $\hat{\mS}$, then for each of the cases \eqref{eq:Xp}--\eqref{eq:Xpm} we have
\begin{align}
\text{Case} \ \eqref{eq:Xp}: \ \sm(\mS)-\sm(\hat{\mS}) & = d_{z}^{+} - d_{y}^{+} - 1 \geq 0 &\quad& \text{by \eqref{eq:ineq1}}, \nonumber \\
\text{Case} \ \eqref{eq:Xm}: \ \sm(\mS)-\sm(\hat{\mS}) & = d_{z}^{-} - d_{x}^{-} - 1 \geq 0 && \text{by \eqref{eq:ineq2}}, \nonumber \\
\text{Case} \ \eqref{eq:Xpm}: \ \sm(\mS)-\sm(\hat{\mS}) & = (d_{z}^{-}-d_{x}^{-}) + (d_{z}^{+}-d_{y}^{+}) - 1 \geq 0 && \text{by \eqref{eq:ineq3}}. \nonumber
\end{align}
Thus, $\sm(\mS)-\sm(\mX) \geq \sm(\hat{\mS})-\sm(\mX)$, showing we can remove the cases \eqref{eq:case1} and \eqref{eq:case4-1}.  

For simplicity, we will now assume that for the original partition $\mS$, all pairings $(x_{i},y_{i})$ are in cases \eqref{eq:case2}, \eqref{eq:case3}, and \eqref{eq:case4-2}--\eqref{eq:case4-4}.  Thus, if we define the sets of pairings
\begin{align}
P^{-} & = \{(x_{i},y_{i}) \ | \ (x_{i},y_{i}) \ \mbox{is in case \eqref{eq:case2}, \eqref{eq:case4-2}, or \eqref{eq:case4-4}}\}, \nonumber \\
P^{+} & = \{(x_{i},y_{i}) \ | \ (x_{i},y_{i}) \ \mbox{is in case \eqref{eq:case3}, \eqref{eq:case4-3}, or \eqref{eq:case4-4}}\}, \nonumber
\end{align}
then we have
\begin{align}
\sm(\mS)-\sm(\mX) & = \Omega^{-} + \Omega^{+} - n \nonumber \\
                  & \geq \quad \sum_{(x_{i},y_{i})\in P^{-}} (d_{y_{i}}^{-} - d_{x_{i}}^{-})
                  + \sum_{(x_{i},y_{i})\in P^{+}} (d_{x_{i}}^{+} - d_{y_{i}}^{+}) - n \nonumber \\
                  & \geq 0, \nonumber
\end{align}
with the last line following from \eqref{eq:ineq1} and \eqref{eq:ineq2}.  The proof is now complete.
\edprf
We immediately have the following corollary, which gives the first degree sequence characterization of split digraphs and is analogous to Theorem~\ref{thm:split1} for split graphs.
\begin{cor}
\label{cor:dschar1}
$d$ is split if and only if there exists $(k,l) \notin \{(0,N),(N,0)\}$ such that $\Sig_{kl} = 0$.
\end{cor}
The next section discusses the relationship between the splittance of a directed graph and graphicality.  The section's main results are Theorem~\ref{thm:displit_slack} and Corollary~\ref{cor:dschar2}, with Corollary~\ref{cor:dschar2} the extension of Theorem~\ref{thm:split2} to split digraphs.

\subsection{Directed splittance and graphicality}
\label{sec:digslack}

Similar to Theorem~\ref{thm:erdos}, the next theorem by Fulkerson gives necessary and sufficient conditions for an integer-pair sequence to be digraphic.
\begin{theorem}[Fulkerson \cite{Fulkerson:1960tv}]
\label{thm:fulkerson}
Let $d$ be an integer-pair sequence with $a$ and $b$ positive and negative lexicographical orderings of $d$, respectively.  Then $d$ is digraphic if and only if $\sum_{i=1}^{N} d^{+}_{i} = \sum_{i=1}^{N} d^{-}_{i}$ and for $k=1,\ldots,N$,
\begin{gather*}
\sum_{i=1}^{k} \min[d_{a_{i}}^{-},k-1] + 
\sum_{i=k+1}^{N}\min[d_{a_{i}}^{-},k] \geq \sum_{i=1}^{k}d_{a_{i}}^{+}\\
(and)\\
\sum_{i=1}^{k} \min[d_{b_{i}}^{+},k-1] + \sum_{i=k+1}^{N}\min[d_{b_{i}}^{+},k] \geq \sum_{i=1}^{k}d_{b_{i}}^{-}.
\end{gather*}
\end{theorem}
Thus, for an integer-pair sequence $d$, we can define the {\bf slack sequences} $\{\bar{s}_{k}\}_{k=0}^{N}$ and $\{\ubar{s}_{k}\}_{k=0}^{N}$ by
\begin{eqnarray*}
\bar{s}_{k} & = & \sum_{i=1}^{k} \min[d_{a_{i}}^{-},k-1] + 
\sum_{i=k+1}^{N}\min[d_{a_{i}}^{-},k] - \sum_{i=1}^{k}d_{a_{i}}^{+}, \\
\ubar{s}_{k} & = & \sum_{i=1}^{k} \min[d_{b_{i}}^{+},k-1] + \sum_{i=k+1}^{N}\min[d_{b_{i}}^{+},k] - \sum_{i=1}^{k}d_{b_{i}}^{-}.
\end{eqnarray*}
Note that $\bar{s}_{0} = \ubar{s}_{0} = 0$ and $\bar{s}_{N} = \ubar{s}_{N} = 0$.

The main theorem of this section is Theorem~\ref{thm:displit_slack}, which shows
\begin{equation}
\label{eq:displit_slack}
\sigma(\vec{G}) =  \min\left\{\bar{s}_{1},\ldots,\bar{s}_{N-1},\ubar{s}_{1},\ldots,\ubar{s}_{N-1}\right\}.
\end{equation}
Thus, $\DG$ is split if and only if $\min\left\{\bar{s}_{1},\ldots,\bar{s}_{N-1},\ubar{s}_{1},\ldots,\ubar{s}_{N-1}\right\} = 0$.  This is in direct contrast to the undirected case, where $s_n = 0$ for any $n$ is not a sufficient condition for a degree sequence to be split.  For example, the degree sequence $d = (4\ 3\ 3\ 3\ 3)$ has slack sequence $s = (0\ 0\ 1\ 2\ 2\ 0)$ with $s_1 = 0$ and $s_m = 2$, where $m=4$ is the corrected Durfee number for $d$.  Therefore, by Theorem~\ref{thm:split2} $d$ is not split.  
However, if we consider the directed extension of $d$, shown again here as
\[
\left(\begin{array}{ccccc} 4 & 3 & 3 & 3 & 3 \\ 4 & 3 & 3 & 3 & 3 \end{array}\right),
\]
with slack sequences $\bar{s} = \ubar{s} = s$, then a digraph $\DG$ with this degree sequence is split since $\min\{\bar{s}_{1},\ldots,\bar{s}_{4},\ubar{s}_{1},\ldots,\ubar{s}_{4}\} = \min\{0,1,2,2\} = 0$.

Both example degree sequences \eqref{eq:ex1} and \eqref{eq:dirext} have the property that
\begin{equation}
\label{eq:slacksplit}
\bar{s}_{k} = \min_{l}\Sig_{kl} \qquad \mbox{and} \qquad \ubar{s}_{l} = \min_{k}\Sig_{kl},
\end{equation}
which is proved in full generality in Corollary~\ref{cor:minsig} and Lemma~\ref{lem:slacksplit}.  Thus, the slack sequences are embedded in the splittance matrix.  The following definition defines {\it maximal sequences} which we show give the precise locations in the splittance matrix where \eqref{eq:slacksplit} is satisfied.
\begin{definition}[Maximal sequences]
\label{def:maxseq}
Suppose $d$ is properly ordered with $a$ and $b$ the corresponding positive and negative lexicographical orderings, respectively.  Define the maximal sequences $\{\mlp\}_{l=0}^{N}$ and $\{\mkm\}_{k=0}^{N}$ such that
\begin{eqnarray*}
\mlp & = & \max\,\{i : d_{a_{i}}^{+} \geq l-1 \ \mbox{and if} \ d_{a_{i}}^{+} = l-1, \ \mbox{then} \ a_{i} \in \mB_{l}\}, \\
\mkm & = & \max\,\{j : d_{b_{j}}^{-} \geq k-1 \ \mbox{and if} \  d_{b_{j}}^{-} = k-1, \ \mbox{then} \ b_{j} \in \mA_{k}\},
\end{eqnarray*}
where $\mA_{k} = \{a_{i}\}_{i=1}^{k}$ and $\mB_{l} = \{b_{i}\}_{i=1}^{l}$.
\end{definition}
The maximal sequences $\{\mlp\}_{l=0}^{N}$ and $\{\mkm\}_{k=0}^{N}$ play a similar role to the corrected Durfee number $m$ in the undirected case, as illustrated in the following lemma.
\begin{lem}
\label{lem:sigord}\

$(i)$ For $k$ fixed, $\Sig_{kl}$ is non-increasing for $0 \leq l \leq \mkm$ and strictly increasing for $\mkm < l \leq N$.

$(ii)$ For $l$ fixed, $\Sig_{kl}$ is non-increasing for $0 \leq k \leq \mlp$ and strictly increasing for $\mlp < k \leq N$.
\end{lem}

\bnprf
We will prove $(i)$, with $(ii)$ following analogously.  Let $k \geq 0$ and $l \geq 1$.  We will keep track of the induced partitions $\mX_{kl} = \{\Xpm_{l},\Xp_{l},\Xm_{l},\X0_{l}\}$ and $\mX_{k,l-1} = \{\Xpm_{l-1},\Xp_{l-1},\Xm_{l-1},\X0_{l-1}\}$.  We have
\begin{eqnarray*}
\Sig_{kl}-\Sig_{k,l-1} & = & kl-|\Xpm_{l}| + \sum_{\{b_{l+1},\ldots,b_{N}\}}d_{i}^{-} - \sum_{\mA_{k}}d_{i}^{+} \\
& & - k(l-1)+|\Xpm_{l-1}| - \sum_{\{b_{l},\ldots,b_{N}\}}d_{i}^{-} + \sum_{\mA_{k}}d_{i}^{+} \\
& = & k -\left[|\Xpm_{l}|-|\Xpm_{l-1}|\right] - d^{-}_{b_{l}}.
\end{eqnarray*}
We have two cases on $x_{b_{l}}$. 
In the first case, $x_{b_{l}} \in \Xp_{l-1}\cap\Xpm_{l}$, which gives $|\Xpm_{l}|-|\Xpm_{l-1}| = 1$.  Thus, if $l\leq \mkm$, we have
\[
\Sig_{kl}-\Sig_{k,l-1} = k-1-d^{-}_{b_{l}} \leq 0.
\]
If $l>\mkm$, then since $x_{b_{l}}\in\Xpm_{l} \Rightarrow b_{l} \in \mA_{k}$, we must have $d^{-}_{b_{l}} < k-1$ and thus $\Sig_{kl}-\Sig_{k,l-1} > 0$.

In the second case, $x_{b_{l}}\in \X0_{l-1}\cap\Xm_{l}$, which gives $|\Xpm_{l}|-|\Xpm_{l-1}| = 0$.  Thus, if $l\leq \mkm$, we have
\[
\Sig_{kl}-\Sig_{k,l-1} = k-d^{-}_{b_{l}} \leq 0,
\]
since $l\leq \mkm$ and $b_{l}\notin\mA_{k}$ implies $d_{b_{l}}^{-} \geq k$.  If $l>\mkm$, then $d^{-}_{b_{l}} \leq k-1 < k$, and thus $\Sig_{kl}-\Sig_{k,l-1} > 0$.
\edprf
The following corollary is immediate.
\begin{cor}
\label{cor:minsig}
\[\min_{l} \Sig_{kl} = \Sig_{k\smkm} \qquad \mbox{and} \qquad \min_{k} \Sig_{kl} = \Sig_{\mlp l}\]
\end{cor}
Finally, we have \eqref{eq:slacksplit} by combining Corollary~\ref{cor:minsig} with the following lemma.
\begin{lem}
\label{lem:slacksplit}
\[\bar{s}_{k} = \Sig_{k\smkm} \qquad \mbox{and} \qquad \ubar{s}_{l} = \Sig_{\mlp l}\]
\end{lem}

\bnprf
We will show $\bar{s}_{k} = \Sig_{k\smkm}$, with the other equality proven in an analogous way.  The equality is trivial for $k=0$ and $k=N$, so suppose $1 \leq k \leq N-1$.  Let $l=\mkm$ and let $\mX_{kl} = \{\Xpm,\Xp,\Xm,\X0\}$ be the partition induced by $(k,l)$.  By the definition of $\mkm$ we have $d_{x}^{-} \geq k-1$ for $x \in \Xpm$, $d_{x}^{-} \geq k$ for $x \in \Xm$, and $d_{x}^{-} \leq k-1$ for $x \in \Xp\cup\X0$.  Thus
\begin{eqnarray*}
\bar{s}_{k} & = & \sum_{i=1}^{k}\min[d_{a_{i}}^{-},k-1] + \sum_{i=k+1}^{N}\min[d_{a_{i}}^{-},k] - \sum_{i=1}^{k}d_{a_{i}}^{+} \\
            & = & \sum_{\Xpm\cup\Xp}\min[d_{x}^{-},k-1] + \sum_{\Xm\cup\X0}\min[d_{x}^{-},k] - \sum_{\Xpm\cup\Xp}d_{x}^{+} \\
            & = & |\Xpm|(k-1) + |\Xm|k + \sum_{\Xp\cup\X0}d_{x}^{-} - \sum_{\Xpm\cup\Xp}d_{x}^{+} \\
            & = & \Sig_{k\smkm}.
\end{eqnarray*}
\edprf
Combining Theorem~\ref{thm:displit_split} with \eqref{eq:slacksplit} nearly gives \eqref{eq:displit_slack}, since we have
\begin{equation}
\label{eq:Csig}
\min\{\bar{s}_{1},\ldots,\bar{s}_{N-1},\ubar{s}_{1},\ldots,\ubar{s}_{N-1}\} = \min_{(k,l)\in C}\Sig_{kl},
\end{equation}
where $C = [0,N] \times [0,N] \setminus \{(0,0),(0,N),(N,0),(N,N)\}$.  Since $\sigma(\DG) = \sum_{(k,l)\notin\{(0,N),(N,0)\}}\Sig_{kl}$,
we will address the remaining two cases $(k,l)=(0,0)$ or $(N,N)$ in the following theorem.
\begin{theorem}
\label{thm:displit_slack}
\[
\sigma(\vec{G}) = \min\left\{\bar{s}_{1},\ldots,\bar{s}_{N-1},\ubar{s}_{1},\ldots,\ubar{s}_{N-1}\right\}
\]
\end{theorem}

\bnprf
If we have $(N,N) \in \arg\min_{(k,l)}\Sig_{kl}$, then $\mlp = \mkm = N$ and we have $d_{a_{N}}^{+} \geq N-1$ and $d_{b_{N}}^{-} \geq N-1$, which implies $(d_{i}^{+},d_{i}^{-}) = (N-1,N-1)$ for all indices $i$.  But this means $\vec{G}$ is a complete directed graph, and thus $\mS = \{\Spm\}$ is a split partition.  Thus, $\hat{\mS} = \{\hSpm,\hS0\}$, where $\hSpm = \Spm-\{x\}$ and $\hS0 = \{x\}$ for any $x \in V$, is also a split partition, showing by \eqref{eq:minsplit} and Corollary~\ref{cor:splitpart} that $\Sig_{N-1,N-1} \leq \sm(\hat{\mS}) = 0$.

For $(0,0) \in \arg\min_{(k,l)}\Sig_{kl}$, Lemma~\ref{lem:sigord} implies $\{(N-1,0),(0,N-1)\} \in \arg\min_{(k,l)}\Sig_{kl}$ as well.  Thus, by Theorem~\ref{thm:displit_split} and \eqref{eq:Csig}, we have
\begin{align}
\sigma(\DG) & = \min_{(k,l)\notin\{(0,N),(N,0)\}}\Sig_{kl} \nonumber \\
            & = \min\{\bar{s}_{1},\ldots,\bar{s}_{N-1},\ubar{s}_{1},\ldots,\ubar{s}_{N-1}\} \nonumber
\end{align}
\edprf
This leads to the second degree sequence characterization of split digraphs using slack sequences.
\begin{cor}
\label{cor:dschar2}
$d$ is split if and only if $\min\left\{\bar{s}_{1},\ldots,\bar{s}_{N-1},\ubar{s}_{1},\ldots,\ubar{s}_{N-1}\right\} = 0$.
\end{cor}

\section*{Acknowledgements}

The author would like to thank the editor and referees for helpful comments and suggestions.

\bibliography{splitDigraphs}
\bibliographystyle{plain}

\end{document}